\documentclass[12pt]{article}
\usepackage{amsmath}
\usepackage{graphicx,psfrag,epsf}
\usepackage{enumerate}
\usepackage[super]{natbib}
\usepackage{url} 

\usepackage{moreverb}
\usepackage{siunitx}
\usepackage{subcaption}
\usepackage{caption}
\usepackage{amsfonts}

\providecommand{\noopsort}[1]{}

\newcommand{\de}{\mathrm{d}}

\newcommand{\blind}{1}

\begin{document}
\graphicspath{{figs/}{figs/}}

\def\spacingset#1{\renewcommand{\baselinestretch}%
{#1}\small\normalsize} \spacingset{1}


\if1\blind
{
  \title{\bf Analysing spatial point patterns in digital pathology: immune cells in high-grade serous ovarian carcinomas}
  \author{Jonatan A. Gonz\'alez
  \hspace{.2cm}\\
    Computer, Electrical and Mathematical Sciences and Engineering (CEMSE),\\
    King Abdullah University of Science and Technology, Jeddah, Saudi Arabia\\ 
        and \\
    Julia Wrobel \\
    Department of Biostatistics and Informatics, \\ 
    Colorado School of Public Health, University of Colorado,\\
    Anschutz Medical Campus, Aurora, Colorado, USA \\
      and \\
    Simon Vandekar \\
    Department of Biostatistics, Vanderbilt University Medical Center,\\
    Nashvill, Tennessee, USA
         and \\
    Paula Moraga \\
    Computer, Electrical and Mathematical Sciences and Engineering (CEMSE),\\
    King Abdullah University of Science and Technology, Jeddah, Saudi Arabia\\ 
    }
  \maketitle
} \fi

\if0\blind
{
  \bigskip
  \bigskip
  \bigskip
  \begin{center}
    {\LARGE\bf Analysing spatial point patterns in digital pathology: immune cells in high-grade serous ovarian carcinomas}
\end{center}
  \medskip
} \fi

\bigskip
\begin{abstract}
Multiplex immunofluorescence (mIF) imaging technology facilitates the study of the tumour microenvironment in cancer patients. Due to the capabilities of this emerging bioimaging technique, it is possible to statistically analyse, for example, the co-varying location and functions of multiple different types of immune cells. Complex spatial relationships between different immune cells have been shown to correlate with patient outcomes and may reveal new pathways for targeted immunotherapy treatments.

This tutorial reviews methods and procedures relating to spatial point patterns for complex data analysis. We consider tissue cells as a realisation of a spatial point process for each patient. We focus on proper functional descriptors for each observation and techniques that allow us to obtain information about inter-patient variation.

Ovarian cancer is the deadliest gynaecological malignancy and can resist chemotherapy treatment effective in cancers. We use a dataset of high-grade serous ovarian cancer samples from 51 patients. We examine the immune cell composition (T cells, B cells, macrophages) within tumours and additional information such as cell classification (tumour or stroma) and other patient clinical characteristics. Our analyses, supported by reproducible software, apply to other digital pathology datasets.
\end{abstract}

\noindent%
\vfill

\newpage
\spacingset{1.45} 
\section{Introduction}
Novel spatial omics assays, including multiplex immunofluorescence imaging, imaging mass cytometry, and spatial transcriptomics, offer the ability to provide simultaneous high-resolution spatial and proteomic or transcriptomic cell-level information. These spatial assays are expected to provide valuable spatial information about interactions between cells and cellular neighbourhoods that inform scientific understanding of how the immune system responds within the tumour microenvironment (TME) across tumour types and stages of cancer development \cite{chen2021differential,johnson2021cancer,elhanani2023spatial,van2022multiplex}. Tumour microenvironments are complex and dynamic ecosystems with various cell types, including cancer and immune cells. The presence and function of immune cells in the tumour microenvironment have been shown to play a critical role in tumour growth, progression, and response to therapy. Image processing techniques combined with spatial statistical methods can answer complex questions about the interactions of cell subtypes within the TME \cite{juliaspatial2022}. For example, recent analyses of multiplex imaging data have been applied to ovarian cancer and indicate that spatial interactions of specific types of immune cells are prognostic of patient survival \cite{Steinhart2021SpatialOvarian}. 

Novel analytic methods have been developed to address analysis goals in characterising spatial relationships between cells. For example, Kim et al.\cite{kim2022unsupervised} describe a novel method for analysing single-cell transcriptomics data using a combination of deep learning and graph theory. This approach allows for identifying cell types and exploring their gene expression patterns, which can be especially useful in spatial analysis applications. Similarly, Zhenghao et al. \cite{chen2020modeling} introduce Spatial-LDA, a topic modelling algorithm that considers the spatial organisation of cells in a tissue sample. This method can infer gene expression patterns and identify cell types with a spatial organisation, providing insight into the spatial heterogeneity of the sample. Zhenzhen et al. \cite{xun2023reconstruction} present a new algorithm for solving the $K$-nearest neighbour problem using locality-sensitive hashing. This algorithm has potential applications in spatial analysis tasks such as computer vision and recommendation systems. Together, these papers highlight the importance of spatial analysis methods in modern biology research.

Many spatial omics methods face theoretical and practical limitations that may affect their application due to implicit simplifying assumptions that are often ignored. The theoretical limitations \cite{baddeley2015spatialR} include the (1) assumption of complete spatial randomness (CSR), one of the most common simplifying assumptions, whereby the cell locations occur within the sampled tissue in a completely random fashion; (2) homogeneity or constant intensity, which means that the expected value of the number of cells in every subset of the sampled tissue remains constant; (3) the assumption of no interaction among cells, i.e., no attraction or repulsion; and (4) the independence between immunity markers (or any other characteristic associated with each cell) and cell locations. Most spatial omics methods do not explicitly define their assumptions, limiting their transparency and reproducibility. In addition, many methods do not permit inference across slides (patients), restricting their utility. These practical limitations could impede the results' accuracy, reliability, and replicability. 

Spatial point processes are convenient and widely used mathematical concepts for analysing event distributions that can be considered as points in a geometric space \cite{moller2004}. A variety of methods and models allow us to understand how points are located in some space and how they interact \cite{illian2008, diggle2013book, baddeley2015spatialR}. These methods study the possible independence between points and consider from complete randomness to aggregation or repulsion conceived from various point generation mechanisms \citep{gelfand2010handbook}. Furthermore, marked point process theory, a particular topic of spatial point processes, can be used to study differences in spatial characteristics between samples with more realistic data models. While these methods were not developed for the analysis of features of the tumour microenvironment, they can be used to provide rigorous statistical inference. These methods can be applied in spatial omics analysis to create accurate models of intercellular processes and their relationship with various pathologies \cite{digitalpathology2014} to understand the spatial organisation of various cells in tissues within the tumour microenvironment (TME) \cite{wilson2021oportunitiesmultiplexdata, wilson2022tumor}.

In this tutorial, we use spatial point process methodologies to analyse the spatial distribution and interactions of immune cells in ovarian cancer tissue samples from several patients collected using multiplex immunofluorescence imaging. We focus specifically on B-cells, macrophages, CD8 T-cells, and CD4 T-cells, immune cell subtypes identified as essential players in the tumour microenvironment of ovarian cancer. Through this investigation, we hope to gain a deeper understanding of the spatial organisation of these immune cell subtypes and their association with patient-level outcomes. Taken as a whole, our tutorial explicitly outlines and tests underlying point process assumptions using a large open-source ovarian cancer dataset.
We provide a principled and reproducible set of analyses for single-cell spatial omics data intended to guide future work in this area. To aid in ease of use, we provide code as supplementary material.

\section{Data}
\subsection{Spatial omics data structure}
Multiplex imaging might have more than 40 channels of markers that bind to specific proteins and can be used to establish cell identification and function. Many datasets, including the one we employ in this work, have fewer markers. After image acquisition, which varies substantially based on the imaging platform, image processing steps are applied before data can be analysed as a point pattern, including autofluorescence adjustment, single-cell segmentation, and cell phenotyping. 
Segmentation identifies individual cells, nuclei, and tissue areas (i.e. tumour and stroma regions) so that analyses can be performed at the single-cell level. Then cells are phenotyped or given a cell-type label based on the expression of different immune and other markers using clustering or gating marker channels \cite{geuenich2021automated,seal2022denVar,zhang2022identification}. Finally, cell marker intensities must be normalised to remove non-biological batch effects across slides or other experimental units \cite{transcriptomicsreview2021, wilson2021oportunitiesmultiplexdata,harris2022multiplexslidetoslidevariation}.

\subsection{Ovarian tumour microenvironment data}\label{sec:data}
Data acquisition and processing details can be found in Jordan et al.\cite{jordanetal2020ovariandata}, and Steinhart et al. \cite{Steinhart2021SpatialOvarian}. The data are available through a public repository on Bioconductor, \cite{VectraPolarisData}. Briefly, 128 cancerous tissue samples from 51 human ovaries were obtained through 5-micron slices of the tissue microarray (TME) stained with specific antibodies for CD8 (T cells, C8/144B, Agilent Technologies), CD68 (macrophages, KP1, Agilent Technologies), cytokeratin (CK, tumour cells, AE1/AE3, Agilent Technologies), CD3 (T cells, LN10, Leica), and CD19 (B cells, BT51E, Leica). 
Immune cells were classified according to the following criteria: B-cell (CD19+), macrophage (CD68+), CD8 T-cell (CD3+ and CD8+), CD4 T-cell (CD3+ and CD8-). Throughout this article, we consider those samples with at least eight cells with each of the four categories of immunity markers in order to have a sufficiently balanced sample to perform the analyses. Figure 1 shows a composite image from our dataset (a) and point patterns of immune cells for four randomly selected subjects (b). 

\begin{figure}[h!tb]
	\centering
	\begin{subfigure}{.49\linewidth}
		\includegraphics[height=3in]{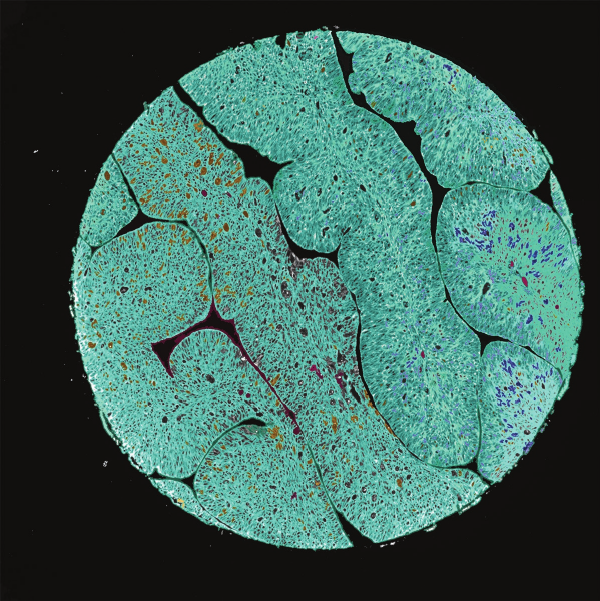}
		\caption{}
	\end{subfigure}
	\begin{subfigure}{.49\linewidth}
		\includegraphics[height=3in]{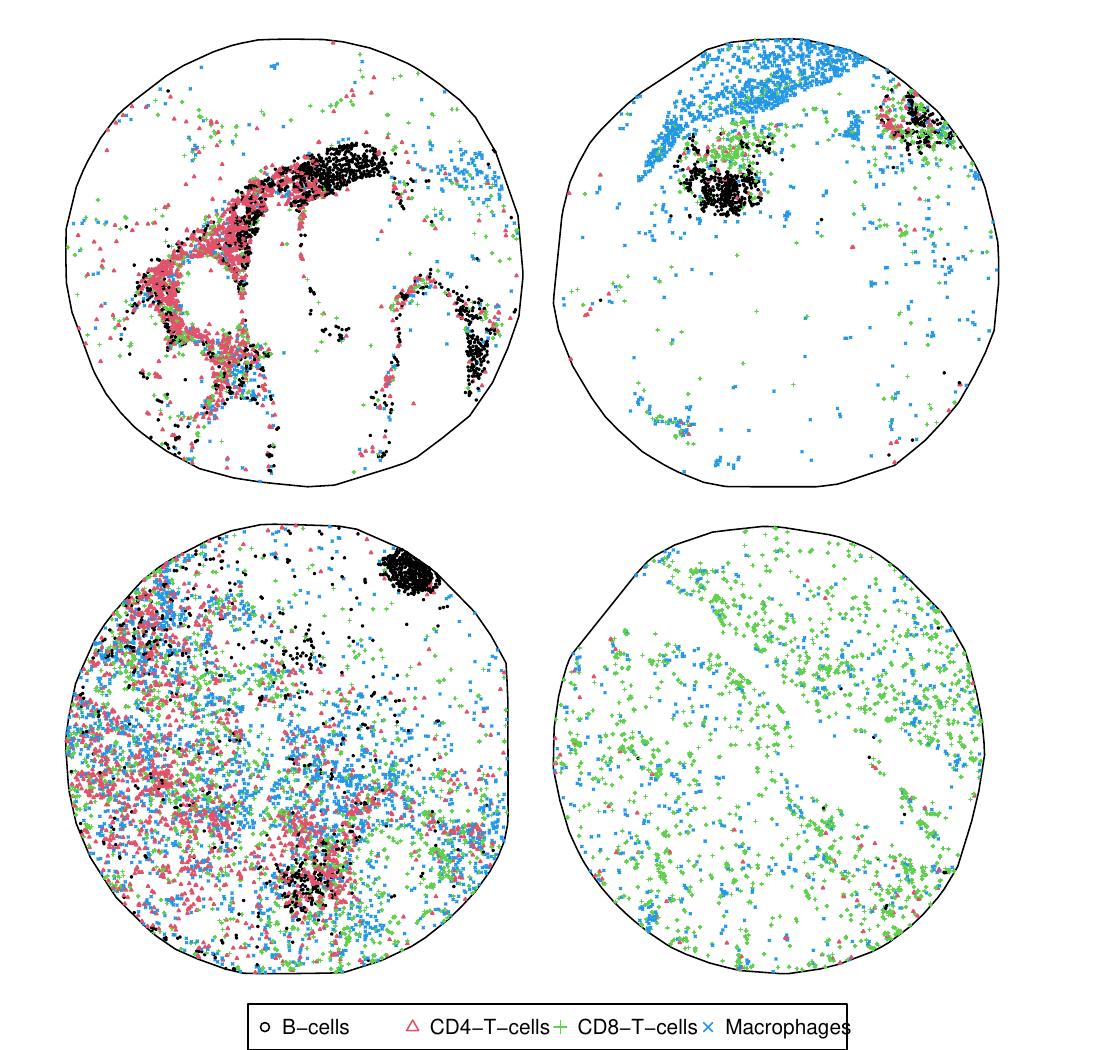}
		\caption{}
	\end{subfigure}
	\caption{(a): Image of an ovarian tumour sample stained with antibodies specific for CD4$^+$ (T-cells), CD8$^+$(T-cells), CD68$^+$ (macrophages), and cytokeratin (tumour cells) from the \texttt{VectraPolarisData} ovarian cancer dataset on Bioconductor. (b): Locations and immune classification of ovarian tissue cells of four out of fifty-one randomly selected patients with high-grade serous ovarian cancer.}
		\label{fig1:cells}
\end{figure}

Each immune cell can belong to the {\it tumour} or {\it stroma} compartment. A tumour is an abnormal growth of cells that can occur in any part of the body. Stroma, on the other hand, refers to the supportive tissue surrounding the cells of an organ or tissue. Additionally, we have several clinical covariates for each patient, including:
\begin{itemize}
    \item Whether the tumour is primary from the initial diagnosis. This is a factor with two categories: Yes or no (binary)
    \item Whether the tumour has undergone chemotherapy before image acquisition (binary)
    \item Whether the patient has a BRCA (breast cancer gene) mutation (binary)
    \item Whether the patient received a \texttt{PARPi} inhibitor (binary)
    \item Cancer stage, a categorical factor which can take the values \texttt{Stage I, Stage II, Stage III, Stage IV}
    \item Patient age (continuous)
    \item Patient survival status (time-to-event)
\end{itemize}

\subsection{Scientific objectives}
The cells in each patient image are considered a multitype marked point pattern. Roughly speaking, a multitype marked point pattern is a collection of points in a spatial domain, where each point is associated with multiple attributes or ``marks'', and the points are categorised into multiple types based on some additional characteristics or attributes. At a high level, our goal is to characterise each patient point pattern and extract features that can be used to analyse patient-level outcomes (e.g., survival). The specific scientific aims we address with this analysis are detailed below.

Our first scientific aim is to explore whether the number of immune cells of different subtypes relates to patient outcomes. This analysis does not incorporate spatial or point process information but is commonly the first analysis step for cell imaging data when several tissue samples (several patients) are available.

Our next scientific aim is to use intensity functions of the point processes, which are first-order point pattern summary statistics described in detail in Section \ref{sec:intensityFunction}, to explore and visualise the density and distribution of cells and continuous-valued functional antibody markers in single images. One quantity we obtain from this analysis is a spatial relative risk of different cell types (e.g., the relative risk of B-cells compared to macrophages in a given tissue sample), and within an image, hypothesis tests can be performed to test the significance of relative risk at a particular spatial location.  Intensity functions of continuous markers, for example, pSTAT3, show how these markers are distributed throughout the tissue. 

The third scientific objective is to use second-order summary statistics of point processes, detailed in Section \ref{sec:SecondOrderDescriptors}, to explain relationships between cell types within the tumour microenvironment. Multitype second-order descriptors such as Ripley's $K$-function explore pairwise relationships between cell types, and we can conduct hypothesis tests to determine if spatial clustering is significant within an image. 

Finally, we discuss how second-order summary statistics of spatial interactions between cells can be associated with patient-level outcomes (i.e., compared across images).

\section{Methods}
\subsection{Point process fundamentals}\label{sec:ppp}
We introduce the following notation: $\mathbf{X}$ will denote a point process and $X$ a point pattern, i.e., a realisation of $\mathbf{X}$ \citep{diggle2003book, baddeley2015spatialR}. The points in a point pattern refer to the spatial locations of a set of individual events or objects that are of interest in a particular study or analysis. In this specific context, points represent cell location within a TME. Therefore, we indistinctly discuss points, spatial or cell locations throughout this paper. \\

$N(B)$ will denote the number of points of $\mathbf{X}$ in a planar region $B\subset W\subset \mathbb{R}^2$. Since spatial locations are recorded with $m$ different labels, one for each cell type, we are interested in multitype point processes analysis. We then divide the process $\mathbf{X}$ into subsets denoted by $\mathbf{X}^{(1)},\ldots \mathbf{X}^{(M)}$, where each $\mathbf{X}^{(m)}$, $m=1,\ldots,M$, is the point process with points of type $m$. In this case, we will denote by $N_m(B)$ the points of type $m$ in the region $B$. When the point process is considered regardless of its labels or {\it marks}, it is denoted by $\mathbf{X}^{(\bullet)}$. Note that the terms ``marks'' and ``markers'' arise from different contexts. The former is related to the characteristics associated with each cell location, for example, cell type or cell size, and it comes from point process analysis, whereas the latter comes from the biological context of antibody markers, for example, B-Cell or macrophages. In order to avoid very complex notation, the subscript of the descriptors will be understood according to the context. The subscripts $j$ and $m$ usually represent marks (B-Cell, macrophages, etc.), and the subscripts $i$ and $k$ usually represent point patterns (tissue samples or patients).\\

In practice, the observations window $W$, which describes the area where cells can occur, is usually attached to the data points. However, when the window is not given, we need to be able to propose a suitable window for the data. This can be done in several ways, for example, by putting all the points inside the minimum rectangle that contains the data or inside the convex hull of the sample. However, these methods have the disadvantage that many data points lie on the edge of the window, which  can introduce unwanted edge effects. A frequently used method that avoids this difficulty is Ripley and Rasson's \cite{ripley1977ripras}, which defines the window as a dilation of the convex hull (centred at the centroid of the convex hull that contains the data) by a factor of $1/\sqrt{1 - \omega / n}$, where $n$ is the number of data points, and $\omega$ is the number of vertices of the convex hull. \\

\subsection{Analysing first-order descriptors}
In univariate statistics, moments are fundamental quantities that provide a valuable mean of describing probability distributions of random variables. The spatial distributions of points have analogous quantities. For example, the numerical means and variances of random variables are replaced by moment measures \cite{daley2007, chiuetal2013stochastic}. Particularly, first-order measures are those associated with the expected values of a variable, which in the case of spatial point patterns, this random variable represents the number of points in each subset of the observation region.

\subsubsection{Intensity function}
\label{sec:intensityFunction}
The intensity function is the first-order descriptor of a point process, i.e., the intensity function describes the expected value of points of some kind anywhere in the observation window; it is given by
\begin{equation}\label{eq:lambdadot}
\lambda_m(\mathbf{u})= \lim_{|\de \mathbf{u}|\rightarrow{0}}\frac{\mathbb{E}\left(N_m(\de \mathbf{u})\right)}{|\de \mathbf{u}|}, \qquad \mathbf{u}\in W, m=1,\ldots,M.
\end{equation}
When $\lambda$ assumes a constant value, the process is called {\it stationary} or {\it homogeneous}.
This assumption is violated in the ovarian cancer data set, as seen by the uneven spatial dispersion of cells in each sample (Figure \ref{fig1:cells}).
The unmarked point process $\mathbf{X}^{(\bullet)}$, which does not distinguish between cell types, has an intensity given by
\begin{equation}\label{eq:lambdabullet}
\lambda_{\bullet}(\mathbf{u}) = \sum_{i=1}^{M}\lambda_i(\mathbf{u}).
\end{equation}

Many techniques are available to estimate first-order intensity functions, including adaptive methods \cite{diggle1985kernel, bithel1990densitykernelepiedemiology, baddeley2015spatialR,Davies2018kernel}. Of all these techniques, non-parametric estimation by kernels is the most common, as well as most appropriate for spatial omics data. Kernel estimation depends on a {\it bandwidth} $(\epsilon)$ that controls the amount of smoothing and is given by the standard deviation of the kernel function. Many methods can be used to estimate the bandwidth: cross-validation \cite{diggle1985kernel,berman1989estimating}, likelihood cross-validation \cite{loader1999local}, Scott’s rule of thumb \cite{scott2015multivariate}, Cronie and van Lieshout’s criterion \cite{cronie2018bandwidth} among others \citep{baddeley2015spatialR}.

Adaptive bandwidth selection is preferred over fixed bandwidth in intensity estimation because it allows a more accurate estimation of the intensity function in different data regions. Depending on the underlying data distribution, fixed bandwidth selection can lead to oversmoothing or undersmoothing of the intensity estimate. On the other hand, adaptive bandwidth selection adjusts the bandwidth locally based on the density of the data points near the location of interest.

In practice, many of these bandwidth estimation methods do not always agree, which often causes the intensity not to be appropriately estimated, especially when some regions of the point pattern are particularly crowded, and some other areas seem to be empty or at least very sparsely populated by points \cite{Davies2018kernel}. This is true in the samples in Figure \ref{fig1:cells}(b). In these cases, assigning a large bandwidth will produce oversmoothing in regions with many points and undersmoothing in areas with few. {\it Adaptive kernel estimators} were developed to address this limitation of fixed bandwidth estimators. We employ an adaptive kernel estimator to accommodate the varying density of cells observed in spatial omics data (Figure \ref{fig1:cells}), defined as follows,
$$
\hat{\lambda}_{m}\left(\mathbf{u}\right) =\frac{1}{e \left(\mathbf{u}\right) }\sum_{i=1}^{n_m}{\kappa_{\epsilon(\mathbf{u}_i)}\left(\mathbf{u}-\mathbf{u}_{i}\right)},\qquad \mathbf{u} \in W, \mathbf{u}_i \in X^{(m)}, m=1,\ldots M,
$$
where $n_m$ is the number of points of $X^{(m)}, \kappa()$ is the Gaussian kernel and $e()$ represents an edge correction factor given by
\begin{equation}\label{eq:intensityedgecorrection}
e \left(\mathbf{u}\right)  = \int_{W} \kappa_{\epsilon(\mathbf{u'})}\left(\mathbf{u}-\mathbf{u'}\right) d \mathbf{u'}.
\end{equation}
The bandwidth function $\epsilon()$ is defined as
$$
\epsilon(\mathbf{u})=\frac{\epsilon^{\star}}{\gamma} \sqrt{\frac{n_m}{\lambda^{\star}_m(\mathbf{u})}},
$$
where $\epsilon^{\star}$ is a fixed {\it global bandwidth} that can be estimated through classical methods such as maximal smoothing principle \citep{terrell1990maximal}; $\lambda^{\star}_j(\mathbf{u})$ is a pilot estimate of the intensity of $X^{(m)}$ usually computed through classical kernel estimation with fixed bandwidth $(\epsilon^{\star})$, and $\gamma$ is the geometric mean term for the pilot intensity evaluated in the points of the point pattern $X^{(m)}$. The geometric mean is intended to free the bandwidth factor from dependence on the data scale \cite{davieshazelton2010adaptiverisk}.

We calculate the global bandwidth using Scott's isotropic rule \cite{scott2015multivariate}. The pilot estimate is the usually fixed bandwidth $(\epsilon^{\star})$ kernel estimate.\\

\subsubsection{Segregation}
In the context of multitype point processes, segregation refers to the tendency of points of different types to spatially separate from one another. A point process is a mathematical model that describes the spatial distribution of points in a region of interest, where each point can be assigned a type or category. Segregation can occur when points of one type tend to cluster together and are spatially separated from points of other types.

In a multitype point process with a spatially varying probability distribution of types, the intensity function $\lambda_m(\mathbf{u})$ can be used to calculate the conditional probability, 
$$
p(m|\mathbf{u}) = \frac{\lambda_m(\mathbf{u})}{\lambda_{\bullet}(\mathbf{u})},
$$ of a point being of type $m$ given its location $\mathbf{u}$. When all the types share a common baseline intensity, i.e., when $\lambda_m(\mathbf{u})=a_m \beta(\mathbf{u})$, $p(m|\mathbf{u})$ becomes constant, meaning that spatial variation in the probability distribution of types is equivalent to segregation of types. To obtain non-parametric estimates of $p(m|\mathbf{u})$, kernel smoothing estimates of the numerator and the denominator can be substituted into the formula.

We can use a non-parametric Monte Carlo test for the spatial segregation between different cell types in a multitype point process \cite{diggle2005segregation}. This test is based on a randomised version of the statistic given by 
\begin{equation*}
T=\sum_{k=1}^n \sum_{m=1}^M \left[ \hat{p}_m(m|\mathbf{u}_k) - n_m/n  \right]^2
\end{equation*}
In this case, $\hat{p}_m(m|\mathbf{u}_k)$ is the leave-one-out kernel estimate of the probability that the $k-$th point has the label $m$.

If the Monte Carlo test indicates that the observed spatial pattern of cells is significantly different from what would be expected under complete spatial randomness, then this suggests spatial segregation between the different cell types. This could imply that the cells are interacting with each other or with the surrounding environment in a non-random manner, leading to the observed spatial organisation.

\subsubsection{Spatial relative risk}
The concept of spatial relative risk dates back to Bithell (1990)\cite{bithel1990densitykernelepiedemiology} and began as a way of viewing and interpreting the relative abundance of cases of a disease with respect to the distribution of the population at risk in a given geographic region. Relative risk is a measure of the strength of association between two densities. It can be used to analyse spatial cell data to compare two cell type densities or, for example, compare a cell type density with the complete density, including all the other types. 

A ratio between two intensity functions is known as {\it relative risk}\cite{kelsalDiggle1995risk}. In general, the spatial log relative risk, defined as a ratio of two intensity functions, is considered in the literature instead of the relative risk and is estimated by 
\begin{equation*}
\rho_{ij}(\mathbf{u})=\log\left\{\frac{\hat{\lambda}_i(\mathbf{u})}{\hat{\lambda}_j(\mathbf{u})} \right\} + \log \left\{\frac{n_j}{n_i} \right\}, \quad i,j =1,\ldots,M,
\end{equation*}

Fixed or variable bandwidth kernels usually perform the numerator and denominator estimators (Section \ref{sec:intensityFunction}). When the bandwidth is fixed, calculating the numerator and denominator bandwidths individually does not usually give good results \cite{davieshazelton2010adaptiverisk}; it tends to under-smooth the relative risk surface. This means that, in general, one would have to choose a considerably larger bandwidth than those used for individual estimates. In the adaptive case, selecting a joint global bandwidth is convenient as doing the pilot estimates separately, i.e., we set a shared adaptive bandwidth in numerator and denominator. Terrel's maximal principle \cite{terrell1990maximal} is often used for estimating the global bandwidth \cite{davieshazelton2010adaptiverisk}.

By examining the relative risks across different regions within the ovaries, we can identify areas with higher or lower chances of a given cell type and potentially identify patterns or clusters of high-risk areas. However, it is essential to remember that the relative risk alone cannot establish causality and that other factors, such as confounding or bias, may also influence the results.

\subsubsection{First-order analysis of continuous marks}
Some marker channels for spatial omics data can be continuously expressed within the tissue. We call ``marks'' these characteristics. For example, among the many marks of the TME of a given tissue sample, our dataset has three: ck marker, ki67 marker, and pSTAT3. 
Specifically, pSTAT3 is a functional marker used to detect a specific type of cell signalling. 

We can consider a spatially weighted average of the marks at each point in the observation window for these markers. This measure can be calculated using a Nadaraya Watson estimator \cite{nadaraya1989nonparametric},
\begin{equation*}
\tilde{m}(\mathbf{u})=\frac{\sum_i m_i \kappa_{\epsilon}(\mathbf{u} - \mathbf{u}_i)e^{-1}(\mathbf{u}_i)} {\sum_i \kappa_{\epsilon}(\mathbf{u} - \mathbf{u}_i)e^{-1}(\mathbf{u}_i)}, \qquad \forall \mathbf{u}\in W,
\end{equation*}
where $m_i$ are the mark values, $\kappa_{\epsilon}()$ is a kernel function with bandwidth $\epsilon$, and $e(\mathbf{u}_i)$ is Eq. \eqref{eq:intensityedgecorrection} evaluated in each data point, this factor is known as Diggle's edge correction \cite{diggle1985kernel,baddeley2015spatialR}.

Through this technique, we can understand the distribution of specific continuous variables within the observation region. Variables sampled across cell locations can be interpolated to the entire observation region, in this case, the tissue sample of a given patient. Note that this technique allows us to interpolate a continuous mark in all window locations, even if no cells exist. Therefore, we predict in a specific location where it might be impossible for a cell to locate itself.

\subsection{Analysing second-order descriptors}\label{sec:SecondOrderDescriptors}
Just as the first-order intensity is the statistic analogous to the mean in a univariate distribution, second-order statistics are analogous to the variance or covariance of a marked point process. In this case, this measure can be associated with the count of pairs of points and is an index of the statistical association between points \cite{baddeley2015spatialR}. This can rigorously quantify relationships between cell types or markers, such as the spatial association between B-cells and Macrophages in ovarian cancer samples. Several statistics allow for analysing of these second-order statistics. We employ some non-parametric descriptors below.

\subsubsection{Multivariate second-order characteristics}\label{sec:kfunctiontheo}
In the context of spatial point processes, it is common to use second-order summary descriptors, such as the pair correlation function or the $K$-function, to describe the spatial arrangement of points in a point pattern. Homogeneous versions of these descriptors assume that the intensity of points is constant throughout each sample, which may not hold true in the case of spatial omics data. Therefore, it is necessary to use inhomogeneous versions of these descriptors that consider the variations in the point process intensity across different types of points and locations. This approach allows for a more accurate description of the spatial patterns and relationships within the point pattern data.

One of the most popular second-order descriptors for analysing point patterns is Ripley's $K$-function \cite{ripley1977}. Ripley's $K$-function in its inhomogeneous multivariate version can be written by
\begin{equation}
{K}_{ij}(r)=\mathbb{E}\left[\left. \sum_{\mathbf{u}_k \in X^{(j)}}\frac{1}{\lambda(\mathbf{u}_k)}\mathbf{1}\{||\mathbf{u}-\mathbf{u}_k||\leq r\}  \right| \mathbf{u}\in \mathbf{X}^{(i)}\right],
\end{equation}
as long as this value does not depend on the choice of the location $\mathbf{u}$ \cite{moller2004,baddeley2015spatialR} and where $r$ belongs to a suitably chosen range $T=(0,r_0]$, \cite{ho2006}. An estimator for the $K$-function is given by
\begin{equation}\label{kestimator}
\hat{K}_{ij}(r)=\frac{1}{|W|} \sum_{x_{\ell}\in X^{(i)}}\sum_{x_{k}\in X^{(j)}} \frac{\mathbf{1}\{||\mathbf{u}_k-\mathbf{u}_{\ell}|| \leq r\}e(\mathbf{u}_k, \mathbf{u}_{\ell};r)}{\hat{\lambda}_i(\mathbf{u}_{\ell})\hat{\lambda}_j(\mathbf{u}_k)},
\end{equation}
where $|\cdot|$ denotes, in this particular case, the area of the window $W$, and $e(\mathbf{u}, \mathbf{v};r)$ is an edge-correction weight (see, e.g.\ \cite{ripley1988,baddeley2015spatialR}). The $L$-function is a popular transformation of the $K$-function and it is given by\cite{besag1977} 
$$
L(r) := \sqrt{\frac{K(r)}{\pi}}.
$$
This transformation initially intends to convert the theoretical Poisson (Complete Spatial Randomness) $K$-function given by $K_{CSR}(r) =\pi r^2$, to the straight line $L_{CSR}(r) = r$ making a visual inspection of the plot more straightforward. In the multitype context, this benchmark represents the independence or no association between types $i$ and $j$ without the assumption of CSR.

When there are many types of cells, comparing a particular type with the remaining cell types is convenient\cite{vanliesout1999mutivariatej}. For this, we can define a second-order descriptor in the sense of the one presented in Eq. \eqref{eq:lambdabullet}. The {\it dot $K$-function}, $K_{i \bullet}(r)$, is the expected number of cells of any type lying within a distance $r$ of a point of type $i$, standardised by dividing by the intensity of the unmarked point pattern, i.e., 
\begin{equation}
{K}_{i \bullet}(r)=\mathbb{E}\left[\left. \sum_{\mathbf{u}_k \in X^{(\bullet)}}\frac{1}{\lambda_{\bullet}(\mathbf{u}_k)}\mathbf{1}\{||\mathbf{u}-\mathbf{u}_k||\leq r\}  \right| \mathbf{u}\in \mathbf{X}^{(i)}\right],\quad r\geq 0.
\end{equation}
The corresponding $L_{i\bullet}(r)$ is defined similarly. 

Now, consider the {\it cross-type pair correlation function} defined as the derivative of the $K$ in the following way\cite{illian2008,chiuetal2013stochastic},
$$
g_{ij}(r):=\frac{1}{2 \pi r}\frac{\text{d}}{\text{d}r}K_{ij}(r).
$$
This function considers the contributions of points lying at distances equal to $r$, and its benchmark under CSR (in the univariate case) is 1. In the multivariate case, the benchmark is also one, but it represents independence when the indexes $i$ and $j$ are different. Specifically, $g_{ii}(r)$ is consistent with CSR, and $g_{ij}(r), i\neq j,$ is consistent with a lack of correlation between types $i$ and $j$\cite{baddeley2015spatialR}. A pragmatic and useful interpretation is conceived through probabilities: the probability $p(r)$ of finding two points of types $i$ and $j$ at locations $\mathbf{u}$ and $\mathbf{v}$ separated by a distance $r$ is 
$$
\lambda_i(\mathbf{u})\lambda_j(\mathbf{v})g_{ij}(r)\de \mathbf{u} \de \mathbf{v}, \quad \mathbf{u},\mathbf{v} \in W,
$$
where $\de \mathbf{u}$ and $\mathbf{v}$ stand for the area of two infinitesimal regions around $\mathbf{u}$ and $\mathbf{v}$, respectively. 

Multivariate second-order descriptors are statistical tools that can analyse spatial patterns in multiplex imaging data. Specifically, they can be used to investigate the spatial relationships between several different types of molecules or cellular structures in the same tumour microenvironment by measuring the extent to which two types of cells are spatially clustered or dispersed relative to each other. The functions generate plots showing the expected distance between cell pairs of each type as a function of the distance between them, allowing researchers to visualise the degree of clustering or dispersion.

\subsubsection{Shortest distance}
As Baddeley et al. \cite{baddeley2015spatialR} pointed out, there is sometimes a confusing ambivalence between counting points in a point pattern in an observation region and measuring the shortest distances. For instance, we can say that there are four cells (on average) in a given \SI{1}{\micro\metre} tissue sample or that one cell appears every \SI{0.25}{\micro\metre}. This duality impacts the mathematics behind point pattern analysis; therefore, a complete statistical analysis should include the study of shortest distances in addition to correlation.

Let $\mathbf{X}$ be a spatial stationary multitype point process and $\mathbf{u}\in \mathbb{R}^2$ a fixed location. Then if we consider the distance $d(\mathbf{u}, \mathbf{X}^{(m)}):=\min\left\{||\mathbf{u}-\mathbf{u}_k||:\mathbf{u}_k \in \mathbf{X}^{(m)}\right\}$ (called the {\it empty-space distance}), the {\it empty-space function} is the cumulative distribution function of the empty-space distance, i.e., 
$$
F_m(r):=\mathbb{P}\left\{d(\mathbf{u},\mathbf{X}^{(m)})\leq r\right\},
$$
for all the interpoint distances $r$. The {\it nearest neighbour distance distribution function} $G_{ij}(r)$ is the cumulative distribution function of the nearest neighbour distance from a point of type $i$ to the nearest point of type $j$, which can be written by as $d(\mathbf{u}, \mathbf{X}^{(j)} \setminus \mathbf{u})$, i.e., the shortest distance from a point $\mathbf{u}$ to the process $\mathbf{X}^{(j)}\setminus \mathbf{u}$. Thus,
$$
G_{ij}(r):=\mathbb{P}\left\{d(\mathbf{u}, \mathbf{X}^{(j)} \setminus \mathbf{u})\leq r | \text{ }\mathbf{u}\text{ is a typical point of }\mathbf{X}^{(i)} \right\},
$$
for any location $\mathbf{u}$. The theoretical values of the univariate versions of these functions for CSR processes are $F_{\text{CSR}}(r)=G_{\text{CSR}}(r)= 1-\exp\{-\lambda \pi r^2\}$, where $\lambda$ is the intensity of the point process; it means that for completely random patterns, the empty space and the nearest-neighbour distance have the same distribution \cite{diggle2003book,moller2004,baddeley2015spatialR}.  In the multitype case, $G_{ij}$ measures the association between types $i$ and $j$, when there is independence, $G_{ij}(r)=F_{j}(r)$.

In practice, it is convenient to compare these two functions ($F$ and $G$) because although in the case of independence (or CSR in the univariate case), they coincide, in the other cases, they tend to behave oppositely; that is, while one increases, the other decreases. We can then consider a new function as the quotient between these two, simultaneously summarising both behaviours. This quotients is known as the $J$-function, defined as \cite{vanlieshoutbaddeley1996,Cronie2015}
$$
J_{ij}(r):=\frac{1-G_{ij}(r)}{1-F_{j}(r)}, \quad \forall r\geq0,\text{ such that }F_{j}(r)<1.
$$
Under independence, $J_{ij}(r)=1, \forall r$, values of $J$ above 1 represent repulsion and values below 1 represent attraction.

In the multivariate non-stationary case, the functions $F_{j}$, $G_{ij}$ and $J_{ij}$ require advanced mathematical treatment to be defined and estimated\cite{Cronie2015}. However, its interpretation remains as simple as in the stationary bivariate case. This is why we recommend its use as a complement to other descriptors such as $K$ and $G$. In the same vein, it is possible to define dot versions, $G_{i\bullet}$ and $J_{i\bullet}$ by considering distances to points of any other type in a multitype point process.

\subsubsection{Complete Spatial Randomness and Independence}\label{sec:CSRI}
When the data points are of a single type, we ignore cell type; the null reference model is complete spatial randomness \cite{diggle2013book,baddeley2015spatialR}. Although this model is unrealistic for most real-world phenomena, it functions as a {\it dividing hypothesis}\cite{cox1977significance} or benchmark from which spatial aggregation or inhibition (regularity) is inferred.

For point patterns of various types, there is an analogous reference model to which complete spatial independence is added to complete spatial randomness. For multitype point patterns, there are two possible choices for a benchmark: random labelling, where type labels are randomly assigned to points, and independence, where points of different types are independent. When both conditions hold together, the point process is known as having {\it Complete Spatial Randomness and Independence} (CSRI).

In the homogeneous case, the theoretical values of all second-order descriptors such as $K_{ij}$, $L_{ij}$, $g_{ij}$, $K_{ii}$, $L_{ii}$, $g_{ii}$, $K_{i\bullet}$, $L_{i\bullet}$, $g_{i\bullet}$, $G_{ij}$, $J_{ij}$, $G_{ii}$, $J_{ii}$ and $F_i$ are consistent with CSRI. Therefore, performing statistical tests to detect discrepancies is easy. However, this does not necessarily occur in the inhomogeneous case. For this case, the $J$-function can be used to test independence and random labelling.

The traditional approach to test independence between two types of points $i$ and $j$ consists of selecting one of the point patterns, e.g., $X^{(i)}$, and making random displacements of its points as many times as desired\cite{lotwick:silverman:82}. Each shifting breaks the possible spatial dependency between types without affecting the dependency within the type. After each displacement, the desired second-order descriptor is calculated and stored in a vector of observations on which a global envelope test can be performed later (see Section \ref{sec:GET}). Each random shifting will produce some points that lie outside the reference window. There are some methods for dealing with these points \cite{tomas2020revisiting}; 
in our case, we use the erosion technique\cite{tomas2020revisiting}. This method uses an eroded window $W_c$ such that $W_c\setminus \mathbf{u}\subset W$ for all possible shift vectors $\mathbf{u}$. Then the statistics are computed with the restricted point patterns to the new window, i.e., $X^{(i)}|_{W_c}$. Since some points are lost when eroding, this method often loses power in the tests. Shifting a point pattern $X$ could affect its distribution in the inhomogeneous case since the intensity changes with the observation region. The intensity of the pattern should be shifted with each random displacement to fix this problem.

In the case of random labelling, the null hypothesis is that each cell immunity marker (the label) is determined randomly, independently of other cells, with constant probabilities. A Monte Carlo test may be deliberated in the following way: To generate different observations of the wanted second-order statistic, it is only necessary to randomly permute the labels through the cells of the tissue samples without changing their locations. If the null hypothesis is true, e.g., if the immunity markers are random, then the point patterns that result from the relabelling are statistically equivalent to the original point pattern. The ``dot'' functions are the most useful for evaluating random labelling for envelope tests\cite{baddeley2015spatialR,Cronie2015}. 

\subsubsection{Global envelope test} \label{sec:GET}
The statistical treatment of point patterns is usually complex, especially when testing hypotheses about the interaction (covariance) between points. Summary statistics are often used to depict the characteristics of the observations in a mathematical fashion; for example, various statistics capture spatial interaction. These functions depend on interpoint distances $r$; we denote this type of function as $T(r)$. We described classic descriptors here: Ripley's $K$ function, the $L$ function, the nearest neighbour distance distribution function $G(r)$, the empty space function $F(r)$ and the $J$ function \cite{moller2004, illian2008, chiuetal2013stochastic, diggle2003book, baddeley2015spatialR}.

Choosing a proper second-order functional descriptor involves assessing the research question and the characteristics of the point pattern data. For example, the cross-type $K$-function helps to identify the clustering or regularity of points of different types relative to each other, while the $J$-function is useful in detecting the association between different marks. It is essential to select the appropriate second-order functional descriptor for the research question and data properties, considering the types of marks, spatial scales of interest, and spatial dependence \cite{baddeley2015spatialR}.

Once we have chosen an appropriate statistic that summarises a point pattern, testing against a null model is helpful to determine whether the cell types or markers exhibit interesting spatial patterns. We want to know if our estimated statistic from the data is incompatible with its distribution under some null model. In practice, the null distribution is often unknown, so using Monte Carlo techniques to generate the null distribution is typical in spatial statistics. Once the necessary simulations are done, the {\it envelope tests} are performed. 

In the context of multiplex imaging, functional envelope tests can be applied to investigate the distribution of any functional descriptor or biomarker calculated from the TME data. For example, researchers may be interested in studying the correlation between the expression of two proteins in the same tissue sample and how this correlation varies across different tissue regions. Envelope tests can provide a way to quantify the degree of similarity or difference between functions computed from the samples and functions computed from synthetic data.

Normally, we can simulate point patterns from some null model. More precisely, we can generate simulations under $H_0:T(r)=T_{\text{obs}}(r)$. Then we can apply the following steps to test the null hypothesis. 
\begin{enumerate}
    \item Compute the test statistic $T_0(r):=T_{\text{obs}}(r)$, for example Ripley's $K$-function.
    \item Generate $s$ simulated patterns $X_i, i= 1,\ldots,s$ from the null model, for example, Poisson (no interaction).
    \item For each $i = 1,\ldots,s,$ compute the test statistic $T_i(r)$ based on the point pattern $X_i$.
\end{enumerate}
We can use a non-parametric global envelope test based on a measure called {\it the extreme rank length} (ERL) \cite{tomasute2017,tomas2020ANOVA}. The ERL measure can compare the functions $T_i$ without multiple testing problems \cite{tomasute2017multiple}. The test may also be interpreted graphically as it points out the distances where the data contradicts the null hypothesis. 

The functions $T_i$ should be evaluated on a fixed number of distances $r_1,\ldots,r_d\in (0,r_0]$, so every function in finitely discretised by $T_i = T_i(r_1),\ldots, T_i(r_d)$.  The ERL measure ranks the $T_i$ among each other. If $R_{0j}, R_{1j},\ldots, R_{sj}$ are the ranks of $T_0(r_j), T_1(r_j),\ldots, T_s(r_j)$, such that the largest $T_i(r_j)$ has rank 1. The vector of pointwise ranks $(R_{i1}, R_{i2}, \ldots, R_{id})$ is associated with each $T_i$. We define $\mathbf{R}_i:= (R_{i[1]}, R_{i[2]},\ldots, R_{i[d]})$ as these pointwise ranks ordered from smallest
to largest, i.e. $R_{i[j]} \leq R_{i[j']}$ if and only if $j \leq j'$. The ERL measure is then defined by using the lexicographic ordering $\prec$ of the $\mathbf{R}_i$ \cite{tomasute2017}, 
and the ERL measure is given by
$$
E_i=\frac{1}{s+1}\sum_{j=1}^s \mathbf{1}\left\{ \mathbf{R}_j \prec \mathbf{R}_i\right\}.
$$
Then, the $p$-value associated with the Monte Carlo test is 
$$
p=\frac{1}{s+1} \sum_{i=0}^{s} \mathbf{1}\left\{E_i \leq E_1 \right\}.
$$
In the case of composite null hypotheses, where some parameters of the null model must be estimated, the tends to be conservative. There are some double-stage Monte Carlo testing approaches to tackle this issue \citep{daogenton2014montecarlotwostage,baddeley2017twostage}.

\section{Results}

\subsection{Associations between cell counts and phenotypic variables} 
Before getting into any spatial analyses, it is important to understand how the counts of the immune cells are related to continuous and categorical regressors. To analyse possible differences in cell populations between patients, we use a linear log-Poisson model for the expected immune cell counts. We ungroup each patient count into several cell types corresponding to the number of B-cells, CD4 T-cells, CD8 T-cells and Macrophages within the two tissue domains of the ovary samples: stroma or tumour. In this particular context, let $n_{ijk}$ denote the number of immune cells of the patient $i, (i = 1,\ldots, g)$, for cell type $j, (j=1\ldots,m)$ of type of tissue $k, (k=1,\ldots,l)$. In our case, $g=51, m = 4$ and $l=2$. A Poisson log-linear model for the expected cell counts may be used to analyse possible differences in counts, considering the possible overdispersion of the data \citep{mccullagh1989generalized}. 

However, we opt in this context for a {\it Generalised Estimation Equation (GEE) model} \cite{liang1986geemodels}. These statistical models are used to analyse {\it longitudinal data}, which involve multiple measurements of the same individuals or patients. In longitudinal data, observations within each patient might be correlated, and traditional statistical models that assume independent observations are inappropriate. GEE models extend the generalised linear model (GLM) framework to account for such correlation and allow for consistent estimation of model parameters, even when the covariance structure is unknown or misspecified.

We consider the number of points of each tissue sample over the sampling areas as an offset; it is included in the model to account for differences in the tissue sample size and the number of points. We also model the possible effects of design covariates $\mathbf{X}$ on cell counts. The model can be expressed as 
\begin{equation}\label{eq:quasipoissonregression}
    \mathbb{E}\left(n_{ijk}| \mathbf{X}\right)=|W_i|\exp{\left(\mathbf{X}^{\top}\cdot \beta \right)}, 
    \quad \text{and} \quad 
    \mathbb{V}\text{ar}\left(n_{ijk}| \mathbf{X} \right)=\phi \mathbb{E}\left(n_{ijk}| \mathbf{X} \right),
\end{equation}
where $\beta$ is the regression coefficient vector. $\phi$ accounts for potential overdispersion. Table \ref{table:coefficients} shows the summary of the fitted model.
\begin{table}[h!tb]
\begin{center}
\scalebox{.85}{
\begin{tabular}{l l c c c r l}
\hline
Regressor           &               &Estimate    &Na\"ive S.E.  &Robust S.E  &$p$-value       &\\
\hline
Intercept $(\beta_0)$&              &$0.03$      &$0.83$        &$1.58$      &$0.9856$          &\\
Immune              &&&&&&\\
                    &CD4 T-cells    &$-0.07$     &$0.33$        &$0.24$      &$0.7785$        &\\
                    &CD8 T-cells    &$1.56$      &$0.22$        &$0.16$      &$<$\num{2e-23}  &${***}$\\
                    &Macrophages    &$1.43$      &$0.24$        &$0.18$      &$<$\num{2e-15}  &${***}$\\
Tissue              &&&&&&\\
                    &Tumour         &$1.38$      &$0.16$        &$0.13$      &$<$\num{2e-25}  &${***}$\\
Primary tumour      &&&&&&\\
                    &Yes            &$-0.06$     &$0.19$        &$0.29$      &$0.8391$        &\\
Prior chemo         &&&&&&\\
                    &Yes            &$-0.58$     &$0.20$        &$0.33$      &$0.0769$        &${*}$\\
Cancer stage        &&&&&&\\
                    &Stage II       &$-3.18$     &$1.10$        &$2.57$      &$0.2189$        &\\
                    &Stage III      &$-0.68$     &$0.47$        &$0.45$      &$0.1304$        &\\
                    &Stage IV       &$0.07$      &$0.49$        &$0.47$      &$0.8853$        &\\
BRCA mutation       &&&&&&\\
                    &Yes            &$0.49$      &$0.19$        &$0.36$      &$0.1741$        &\\
PARPi inhibitor     &&&&&&\\
                    &Yes            &$0.16$      &$0.17$        &$0.41$      &$0.6990$        &\\
Death               &&&&&&\\
                    &Yes            &$-0.18$     &$0.16$        &$0.29$      &$0.5396$        &\\
\hline
Age at diagnosis    &               &$-0.01$      &$0.01$        &$0.02$      &$0.8003$        &\\
\hline
&&&\multicolumn{4}{r}{\scriptsize{$^{***}p<0.001$; $^{**}p<0.01$; $^{*}p<0.05$}}
\end{tabular}
}
\caption{Estimates with associated na\"ive and robust standard errors for the regression coefficients of generalised estimation equation (gee) model.}
\label{table:coefficients}
\end{center}
\end{table}
The estimated scale parameter is  $\hat{\phi} = 441.7 \gg 1$, indicating an extreme extra-Poisson variation of the cell counts within each combination of the factors considered. This high additional variance could be fundamentally due to a non-Poisson variation (departures from {\it complete spatial randomness} (CSR)) within each sample, for example, due to the attraction or repulsion between cells or to inter-patient variation in mean cell count, possibly related to covariates that could not be measured in the study or random effects.

In Table \ref{table:coefficients}, we see significant increments in the counts of CD8 T-cells, and Macrophages relative to the reference level, B-cells. We also appreciate that there is more abundance of cells in tumour tissue than in the stroma.
A common side effect of chemotherapy is a decrease in immune cell counts, evidenced by chemotherapy patients, who experienced a mildly significant decrease in immune cell counts. 
The other regressors are not statistically significant, meaning there is insufficient evidence for an influence on the counts.

\subsection{Analysing expected counts of immune cells within TMEs}
The first-order descriptors refer to the rate at which immune cells are distributed in the tissue. This function describes the density of immune cells at any given point in the tissue and can be used to understand the spatial arrangements of these cells. 

The observation window $W_i$, where $i$ denotes the patient's number, is defined through Ripley and Rasson's technique \cite{ripley1977ripras}. In this case, we have different spatial windows for different point patterns, i.e., we have different sampling regions for different patients, as every patient has an associated point pattern. In a point process, the intensity function represents the first-order properties, which describes how the mean number of points in a dataset varies with spatial coordinates. This measurement is fundamental in point processes because it shows how the mass or the total number of points is distributed throughout the study region.

In Figure \ref{fig2:intensitiesdiagram}, we show, as an example, a first-order analysis of one of the patients in the sample. We first show the point pattern associated with the patient with the delimitation of the window $(W_6)$ and without differentiating between cell types, that is, without assigning any mark neither discrete nor continuous (Figure \ref{fig2:intensitiesdiagram} (a)). As each cell can have several characteristics, we mainly highlight two of them, the type of immunity and the type of tissue; when we choose one or the other, we will have two different coloured point patterns (Figure \ref{fig2:intensitiesdiagram} (b) and (c)).

\subsubsection{Intensity function}
Specifically, the first-order intensity function provides information about the average number of immune cells per unit area and can help researchers identify regions of the tissue where immune cells are more or less concentrated. Understanding the first-order intensity function of a point pattern in this context can provide insight into the immune response to ovarian cancer and potentially inform strategies for improving immunotherapy treatments.

\begin{figure}[h!tb]
\centering
\includegraphics[width= 1\linewidth]{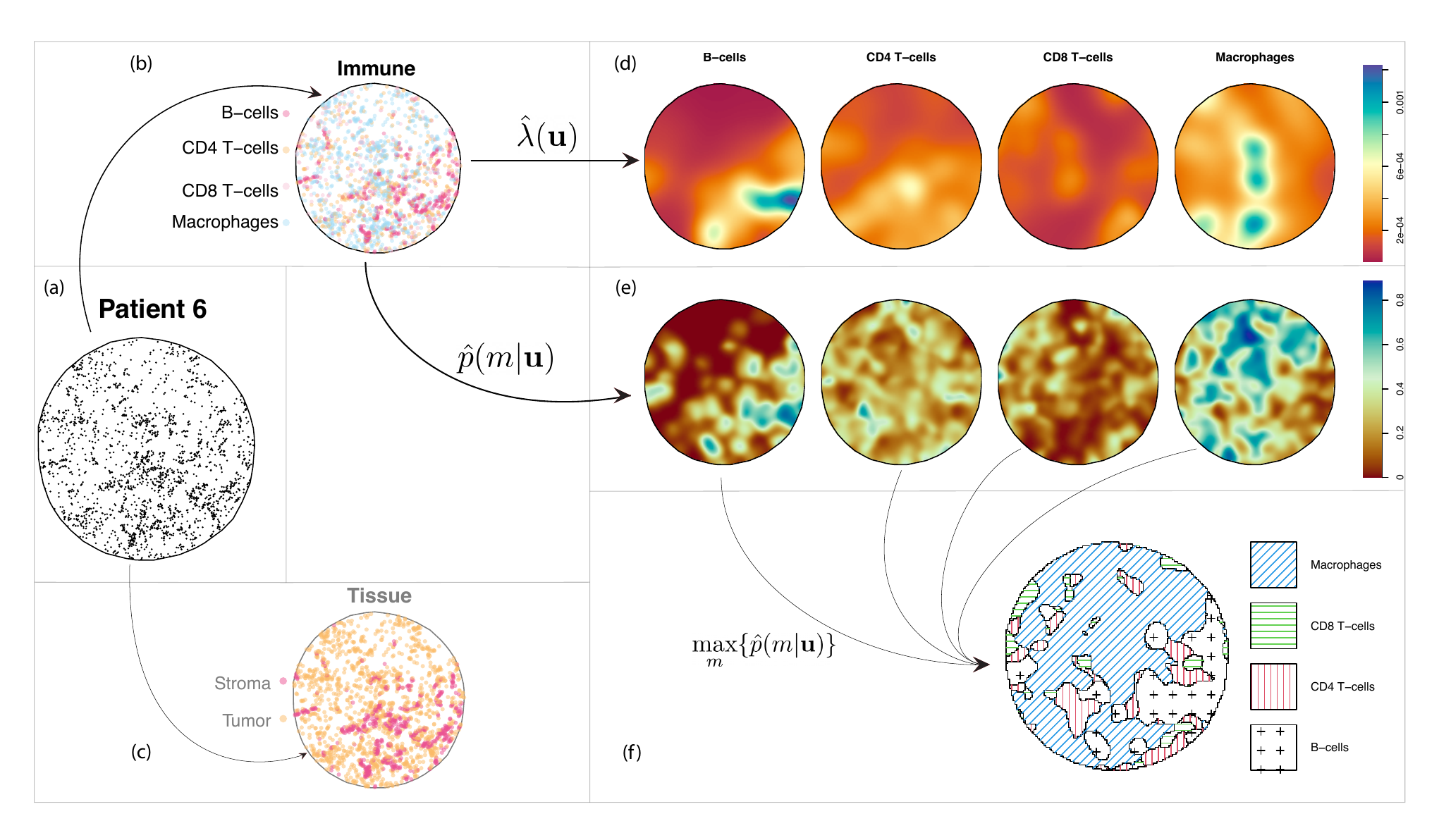}
\caption{Example of kernel estimates of the intensities of patient six. This conveys information on the abundance of the immune cells and the estimated probability of each immunity marker. (a) Representation of the immune cells without considering marks. (b) Representation considering immune markers. (c) Representation considering tissue type. (d) Kernel estimates of intensity for each immune marker. (e) Spatial probability of each immune marker with respect to the other types. (f) Most likely immunity markers at each location of the tissue sample of patient six.}
\label{fig2:intensitiesdiagram}
\end{figure}

Intensity estimates for each immune marker for a particular patient (patient six) are displayed in Figure \ref{fig2:intensitiesdiagram} (d). Note that the estimates can be performed taking into account another set of labels such as tissue region (Figure \ref{fig2:intensitiesdiagram} (c)).
\\

\subsubsection{Segregation}
We can speak about the probability distribution of different cell types. The probability that any cell belongs to type $m$ is 
$$
p_m(m|\mathbf{u}) = \frac{\lambda_m(\mathbf{u})}{\lambda_{\bullet}(\mathbf{u})},
$$
as long as $\lambda_{\bullet}(\mathbf{u})\neq 0,$ where $\lambda_{\bullet}(\cdot)$ is given in Eq. \eqref{eq:lambdabullet}.  Estimates of the varying spatial proportion of each immune marker are displayed in Figure \ref{fig2:intensitiesdiagram} (e), suggesting that immune markers are segregated; for example, Macrophages appear clustered more in the top of the sample, whereas B-cells are shown in higher proportions in the right. Spatial segregation occurs when certain types of points, in particular subregions, predominate rather than randomly mix within some observation window. A texture plot can be used to summarise which immune marker of four has the highest probability at this location, i.e., we divide the tissue sample into regions where different immunity markers predominate (Figure \ref{fig2:intensitiesdiagram} (f)). We can appreciate how macrophages and B-cells predominate throughout patient six's sample.

We can apply this test using, for example, 999 simulations for each region of interest in our dataset. It is typical to adjust the $p$-values to account for multiple tests across the $g$ patients using Bonferroni correction \cite{hochberg1988bonferroni}. We obtain mildly significant $p$-values (about 0.051) for the segregation of immune markers. \\

\subsubsection{Relative risk}
\begin{figure}[h!tb]
\centering
\includegraphics[width= 1\linewidth]{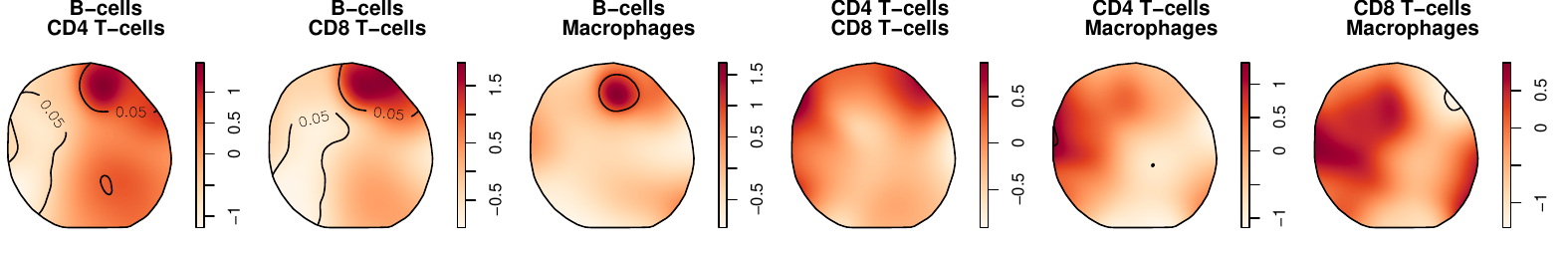}
\caption{Adaptive bandwidth spatial log-relative risk surfaces of patient six data for each combination of levels of Immune markers, with asymptotic tolerance contours.}
\label{fig:relativerisk}
\end{figure}

Understanding the relative risk between different pairs of immune markers in cancerous tissues such as ovaries may be relevant for the development of effective diagnostic and treatment strategies. While each immune cell type plays a distinct role in the immune response against cancer cells, their functions might be intricately connected and can influence one another. By understanding the relative risk associated with each pair of immune markers in cancerous tissues, researchers can identify the most effective combinations of immune markers for targeting cancer cells and develop therapies that enhance the effectiveness of the immune response while minimising the potential negative impact of immune cells that may promote tumour growth. Estimates of spatial varying relative risks of immunity markers are displayed in Figure \ref{fig:relativerisk}.
We can appreciate the northeast region of the tissue sample, particularly with a significantly increased risk (up to 1.5 times) of having B-cells given the counts of CD4 and CD8 T-cells; this risk is also slightly with respect to Macrophages. The southwestern region, on the other hand, shows a significant decrease in the first two cases.

\subsubsection{Smoothing continuous markers}
\begin{figure}[h!tb]
\centering
\includegraphics[width= 1\linewidth]{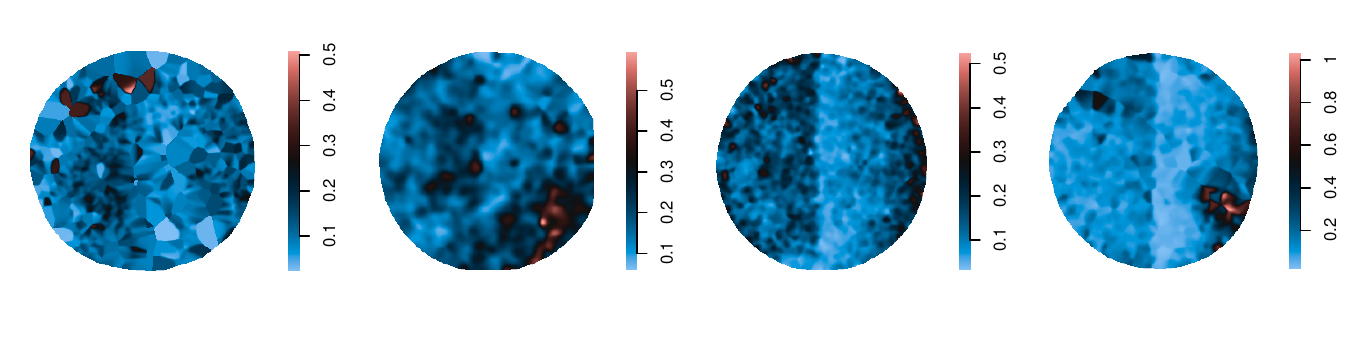}
\caption{Spatially varying average pSTAT3 of four tissue samples (patients).}
\label{fig:smoothingmarks}
\end{figure}

A marker for phosphorylated signal transducer and activator of transcription-3, pSTAT3, is available for all cells in each sample. It is present in multiple types of cells, including tumour cells, immune cells, and stromal cells. pSTAT3 regulates various cellular processes, such as cell growth, proliferation, and survival, and has been implicated in the development and progression of numerous types of cancer. Due to its widespread presence and functional significance, pSTAT3 has become a popular target for cancer therapy research. pSTAT3 is known to be upregulated in ovarian cancer tissue compared to normal ovarian tissue \cite{gao2021impact}. pSTAT3 exhibits significant spatial heterogeneity that differs across patients (Figure \ref{fig:smoothingmarks}).  

\subsection{Analysing interactions amongst cells: correlation and spacing}
At the microscopic scale, neighbour cell-cell interactions in cancer tissue may involve a complex interplay between different cell types. Cancer cells can interact with neighbouring cells through direct contact or by releasing signalling molecules that bind to specific receptors on the surface of neighbouring cells. These interactions can promote cancer cell proliferation, migration, and invasion and modulate the behaviour of surrounding cells, such as immune cells and stromal cells. We approach the complex nature of neighbour cell-cell interactions at the microscopic scale through second-order statistics that depend on a suitable range of distances $r\in T=(0,r_0]$ (see Section \ref{sec:kfunctiontheo}).
In our case, we set a sensible default that depends on the geometry of the ovaries; indeed, $r_0= \SI{350}{\micro\metre}$. 

\subsubsection{Multivariate second-order descriptors}
\begin{figure}[h!tb]
\centering
\includegraphics[width= 1\linewidth]{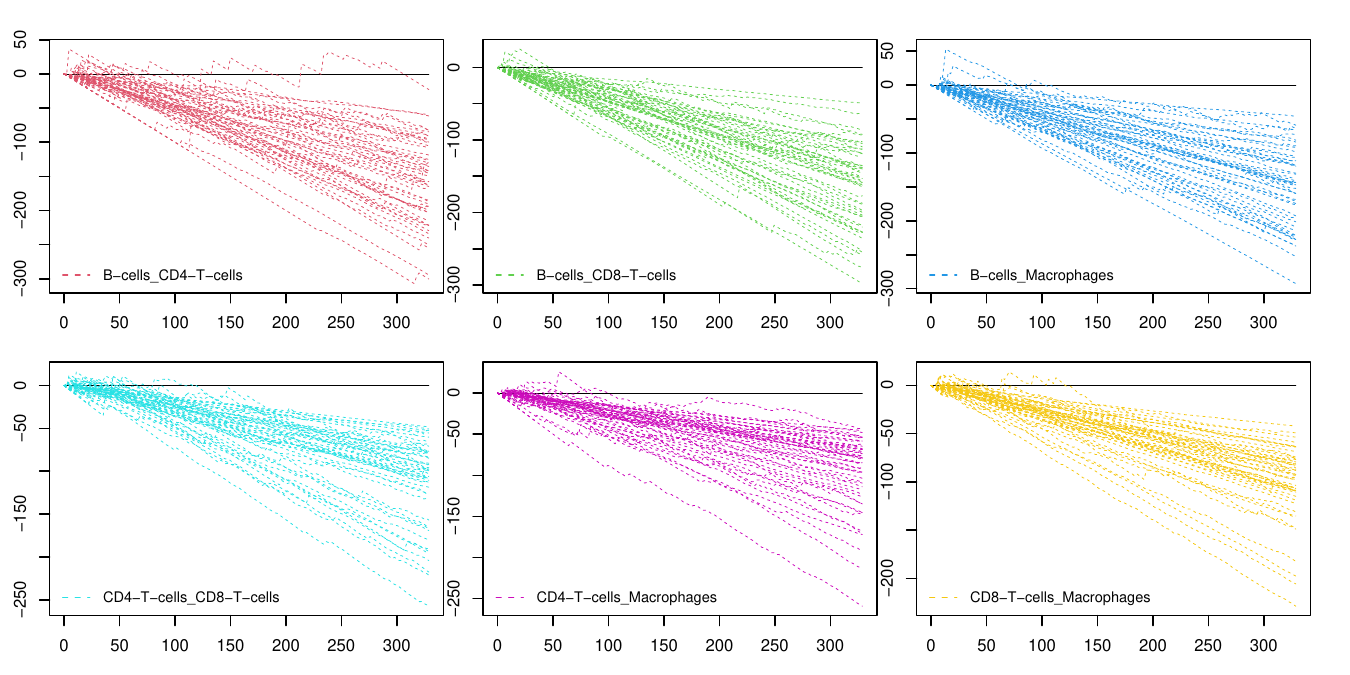}
\caption{The centred cross-type $L$-functions $L_{ij}(r) - r$ for each pair of types $i$ and $j$ of immune cells of all patients. Every colour and panel represents a combination $ij$ of immune types of cells.}
\label{fig:KImmunepercell}
\end{figure}

We use multivariate Ripley's $K$ and pair correlation to quantify associations between cell types across all samples in the dataset. The negative discrepancies between the estimates and $L(r)=0$ (the benchmark) suggest that the cell type pairs show a spatially repulsive behaviour.   
We can arrange the centred $L$-functions by cell type pairs to have a simple judgement about the interaction (Figure \ref{fig:KImmunepercell}).
In this case, where the point pattern represents immune cells of two different immunity markers, these values falling below the independence benchmark on average might suggest that the two types of immune cells are more closely associated with one another than would be expected by chance, i.e., they would interact to each other by inhibiting. This might indicate a situation where the immune system is overwhelmed by the cancer cells, and the immune cells cannot work together efficiently. The inhibitory effect could also be due to the competition between the two types of immune cells for resources or space within the tumour microenvironment. However, this independence hypothesis should be formally tested, and we present that analysis in the next section.

\subsubsection{Complete Spatial Randomness and Independence}
\begin{figure}[h!tb]
\centering
\includegraphics[width= 1\linewidth]{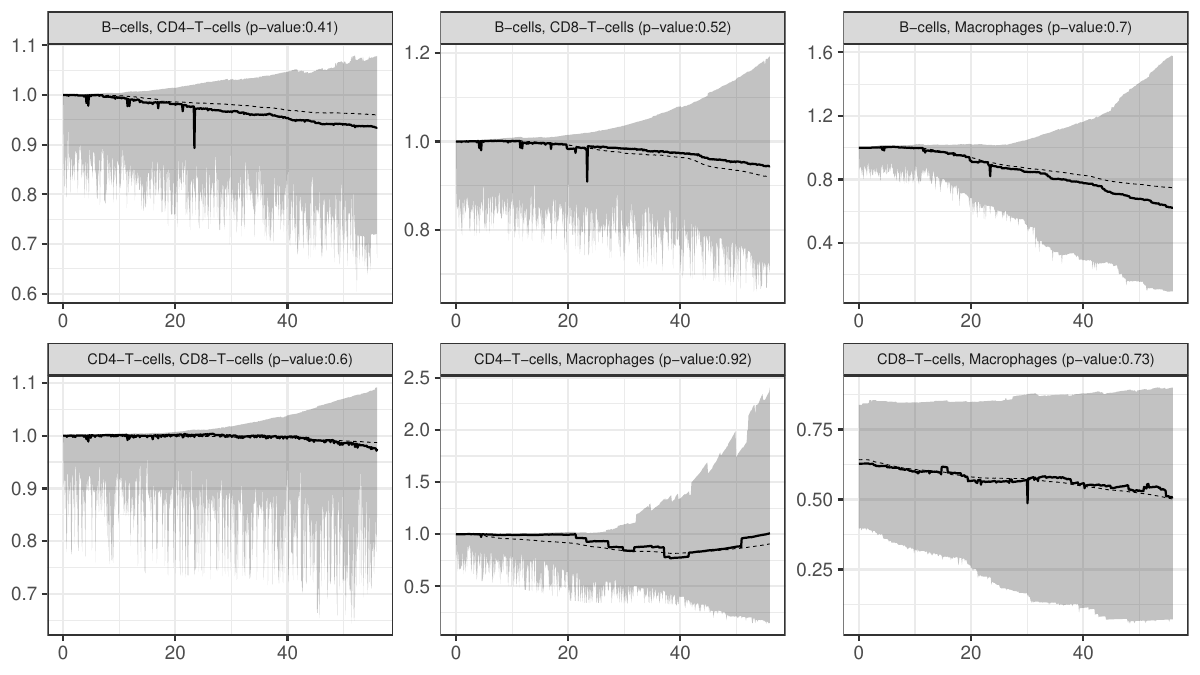}
\caption{$J$-functions $J_{ij}(r)$ for patient 3 where $i$ and $j$ are all the cell types. The envelopes of 2999 simulations were generated by random shifts (grey shading with extreme rank length). Results suggest the different components are independent.}
\label{fig:jindependence}
\end{figure}

To test the independency, we use the bivariate inhomogeneous $J$-function and the shifting technique described in Section \ref{sec:CSRI}. For illustration purposes, we show the test results of a random patient in Figure \ref{fig:jindependence}; the same procedure can be applied to each one in the sample. 
We obtain non-significant $p$-values for every combination of immunity markers indicating that we are in the presence of independence. In the context of immune cells with different immunity markers, the independence of the two types would suggest that the two immunity markers do not have any significant spatial association with each other. This could indicate that the two markers behave independently within the tumour microenvironment. However, it is worth noting that independence does not necessarily mean that the two types of immune cells do not interact. There could still be functional interactions occurring between the two types of immune cells that are not reflected in their spatial distribution. 

\begin{figure}[h!tb]
\centering
\includegraphics[width= .8\linewidth]{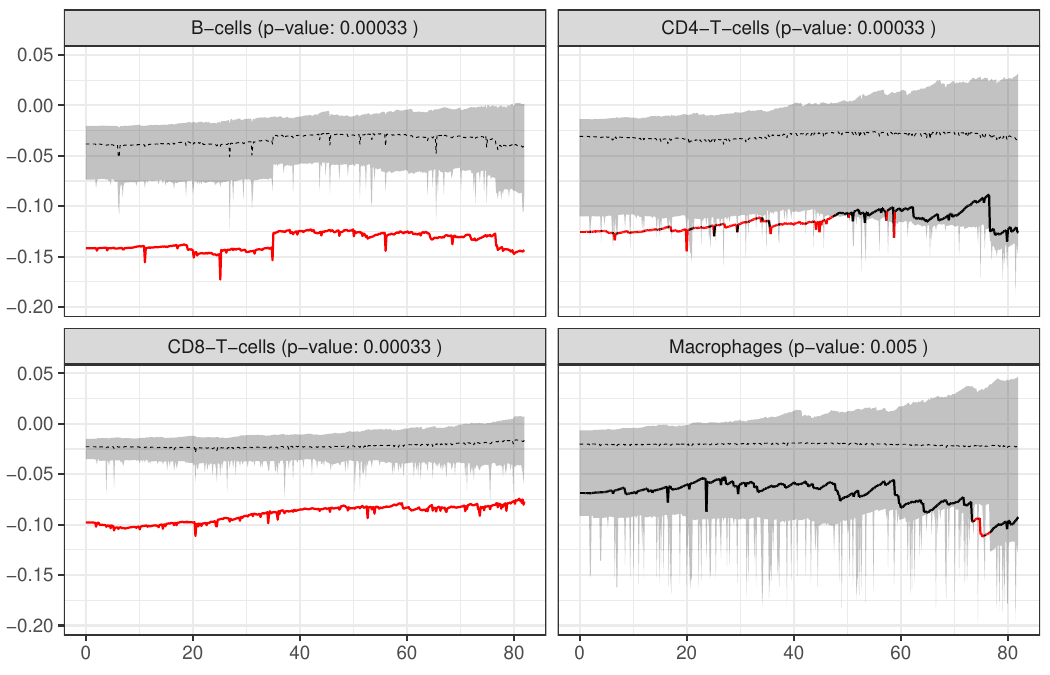}
\caption{Test of random labelling for patient 3. The summary function is $\hat{J}_{i \bullet}(r)-\hat{J}(r)$, where each level $i$ is each panel title.  The solid lines are the row means of 2999 random labellings, and the grey shading represents a central region based on the extreme rank length. Results suggest that the cell types are not random labels.}
\label{fig:jrandomlabelling}
\end{figure}

In the case of random labelling, we use the centred version of the dot $J$-function and a set of 2999 permutations of the labels of the immunity markers. The results for patient 3 are displayed in Figure \ref{fig:jrandomlabelling}.
We obtain enough statistical evidence to reject the random labelling hypothesis in this case. If the immune cells in a cancerous tissue sample have non-randomly assigned immunity markers, it might suggest that there is a selective pressure on the immune cells to express specific markers that enable them to recognise and target cancer cells more effectively, for instance. This selective pressure could be due to the cancer cells, which may have developed mechanisms to evade the immune system and require a more specific immune response. Alternatively, it could be due to external factors such as treatment, which may have induced a selection pressure on the immune cells to express markers more effectively in combating the cancer cells.

\subsection{Testing differences between patients groups}
Finally, we can also test the hypothesis that two (or more) observed point patterns are realisations of the same spatial point process model to investigate whether all types of immune cells interact similarly across patient groups. For example, we would want to compare whether the patients with four different cancer stages have the same spatial distribution of immune cells. We use the inhomogeneous $L$-function in this case, but we could use any other second-order descriptor. The null hypothesis that we would like to test can be formally formulated as,
\begin{equation}\label{eq:H0Lfunctions}
    H_0: \bar{L}_{\text{stage I}}(r) =\bar{L}_{\text{stage II}}(r)=\bar{L}_{\text{stage III}}(r)= \bar{L}_{\text{stage IV}}(r), \quad \forall r\in T.
\end{equation}
We can establish analogous hypotheses for the patients grouped by other factors such as prior chemotherapy or death.

\begin{figure}[h!tb]
\centering
\begin{subfigure}{.49\linewidth}
\centering
\includegraphics[height=2.65in]{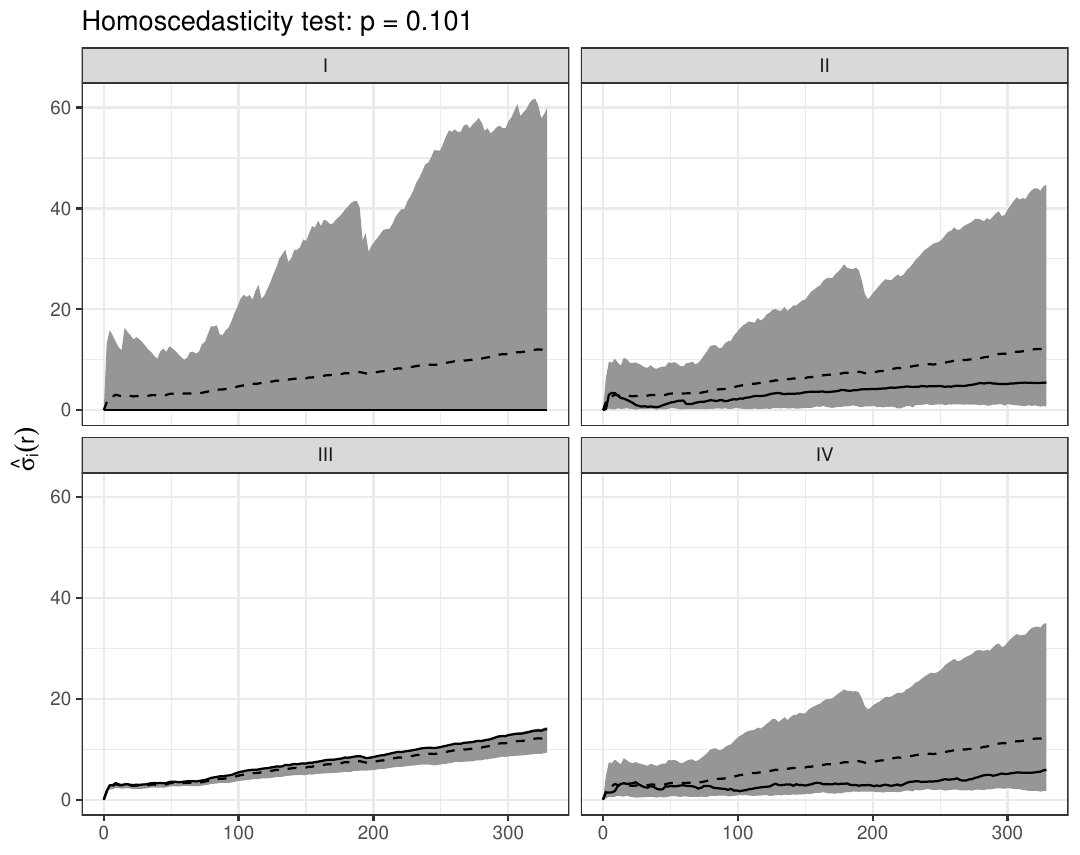}
\caption{}
\end{subfigure}
\begin{subfigure}{.49\linewidth}
\centering
\includegraphics[height=2.65in]{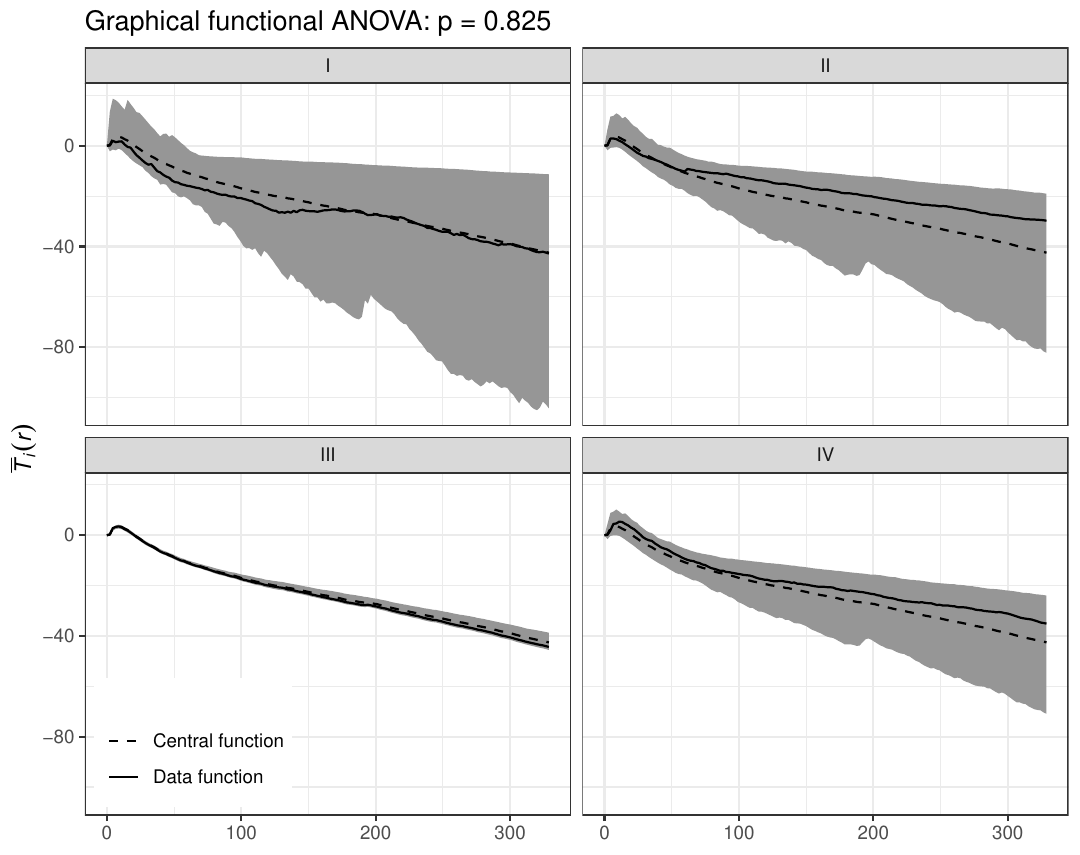}
\caption{}
\end{subfigure}
\caption{(a) The test for equality of variances of the centred $L$-functions in the four cancer stages groups. (b) The graphical functional ANOVA based on functional descriptors. Test for equality of means of the centred $L$-functions in the four groups according to cancer status using the group means.}
\label{fig:kanova1}
\end{figure}

Before using a specific statistic for testing $H_0$, we should evaluate the assumption of homoscedasticity, i.e., that the variances are equivalent across the different patient groups. We can evaluate the equality of variances of the $L$-functions calculated for two or more groups through a Levene's-style test \cite{tomas2020ANOVA}. This equality would be convenient since some statistical procedures assume that the populations' variances are equal. We obtain $p$-values of $0.121, 0.520$ and $0.067$ for cancer stage, prior chemo and death, respectively. It means that the $L$-functions show equal variances across groups. The graphical output for the test for cancer stage as grouping factor is shown in  Figure \ref{fig:kanova1}(a).

To test the hypothesis $H_0$ shown in Eq. \eqref{eq:H0Lfunctions}, we can use the original $L$-functions or a suitable rescaled statistic in case of heteroscedasticity \cite{tomas2020ANOVA}.
We can apply a rank envelope test by permuting the $L$-functions in a non-parametric one-way ANOVA fashion, i.e., permuting the functional descriptors across the different patient groups. We set 50,000 random permutations and obtain  $p$-values of $0.829,0.652$ and $0.225$ for cancer stage, prior chemo and death, respectively. We show the graphical output for the test in the case of cancer stage in Figure \ref{fig:kanova1}(b) for illustration. 
This analysis suggests that the interaction of different immune cells remains consistent across various cancer stages. The presence or absence of chemotherapy did not significantly affect the interactions between these immune cells. Additionally, the analysis found that patients who died during the sampling did not display any statistical differences in cell interaction compared to those who survived. These findings suggest that the cancer stage does not strongly influence the interactions between immune cells, the use of chemotherapy, or the patient's survival. 

\section{Conclusion}
This paper introduces non-stationary spatial point processes to model interactions between cell types in ovarian cancer samples. We have made a tour of the methods of point processes related to the descriptive analysis of observations. These observations or realisations can be analysed separately or together through replicated point process methodologies. We have seen how both first-order characteristics, related to the spatial distribution of the number of points, and second-order characteristics, related to the covariance structure between points, can be estimated. These methods can reveal biologically interesting features of ovarian cancer samples that can inform our understanding of the immune response in ovarian cancer. 

The techniques for analysing replicated patterns help study cell arrangements in multiplex imaging because more than one sample is available, and inference across samples is of primary interest in understanding how spatial features relate to patient-level phenotypes.

Although the number of statistical procedures related to point patterns is vast\cite{illian2008, diggle2013book,baddeley2015spatialR}, the concepts and contexts do not usually fit the reality of cell patterns, especially when the tissue that has been observed is not complete, that is when we observe only a piece of the tissue. This is the case of the ovarian cancer samples analysed here, which present irregular geometric shapes that change from one patient to another. In technical terms, these cell characteristics translate into inhomogeneity and heteroskedasticity in the functional descriptors\cite{chiuetal2013stochastic} since the functional form of the variances usually implies window geometry and intensity. The methods we proposed for multiplex image analysis here are appropriate for inhomogeneous and heteroskedastic point processes, so they are ideal for modelling spatial interactions of cells in these datasets. However, this problem of non-stationarity, far from being a limitation, constitutes a marvellous open field of investigation.

Spatial statistics research in digital pathology has the potential to  enhance our understanding of disease progression and the effectiveness of treatments. However, analysing large-scale digitised tissue images requires powerful computational resources and efficient algorithms to handle the vast amount of data generated. Rapid and scalable computation is essential in this context. It enables researchers to analyse massive datasets in real-time and identify patterns and trends that would otherwise be missed. This is particularly important in digital pathology, where the timely analysis of tissue images can significantly impact patient outcomes. Possible directions for spatial statistics research in digital pathology include developing novel algorithms designed explicitly for high-dimensional data and exploring distributed computing architectures that can leverage cloud-based resources for efficient data processing. Ultimately, the ability to rapidly and efficiently analyse spatial data in digital pathology can greatly enhance our ability to diagnose and treat diseases, leading to better patient outcomes and improved public health.

\bibliographystyle{chicago}

\bibliography{bibliography}
\end{document}